\documentclass[twocolumn,showpacs,preprintnumbers,amsmath,amssymb]{revtex4-1}
\usepackage{epsfig}

\begin{document}

\title{Inelastic cross sections and continuum transitions illustrated by $^{8}$Be results}

\author{E. Garrido$\:^1$, A.S. Jensen$\:^2$, D.V. Fedorov$\:^2$}
\affiliation{$^1$ Instituto de Estructura de la Materia, IEM-CSIC,
Serrano 123, E-28006 Madrid, Spain}
\affiliation{$^2$ Department of Physics and Astronomy, Aarhus University, 
DK-8000 Aarhus C, Denmark} 

\date{\today}

\begin{abstract}
We use the two-alpha cluster model to describe the properties of
$^{8}$Be. The $E2$-transitions in a two-body continuum can be
described as bremsstrahlung in an inelastic scattering process.  We
compute cross sections as functions of initial energy for the possible
$E2$-transitions from initial angular momenta $(0^+,2^+,4^+,6^+,8^+)$.
The dependence on the exact shape of potentials is very small when the
low-energy scattering phase-shifts are the same. We relate to
practical observables where energies of the emerging alpha-particles
are restricted in various ways. The unphysical infrared contribution
is removed.  We find pronounced peaking for photon energies matching
resonance positions.  Contributions from intra-band transitions are
rather small although substantial (and even dominating) for initial energies
between resonances.  Structure information are derived but both ${\cal
  B}^{(E2)}$ values and electromagnetic transition rates are ambiguous
in the continuum.
\end{abstract}


\maketitle

\section{Introduction}

Nuclear structure information is often obtained from reaction
experiments.  Elastic scattering is the conceptually simplest process
which probes the interaction between the colliding particles.  The
same particles in initial and final states, but different relative
energies, is an inelastic process where the missing energy must be in
emitted photons (provided that the particles themselves remain non-excited).  
Such cross section measurements then carry information about continuum
properties of the combined system of the two particles, and
corresponding cluster properties are perhaps here especially
pronounced.  Clearly, structure and dynamics are entangled, and
extraction of useful structure information is then a challenge.

More and more experiments probing continuum properties are in the
pipeline.  Cluster structures are known to be of extreme importance in
light nuclei and in particular for astrophysical processes in the
continuum.  A number of models employ spatially limited sets of basis
states where the continuum has disappeared.  They provide directly
structure information but such discretization is not necessarily
leading to correct observable results.  The dynamics of a given
reaction process has been removed by assumptions of independence of
decay process and initial and final scattering states.

Two-body scattering processes are completely determined by the phase
shifts, which in principle can be reproduced by different potentials.  This is
especially obvious for low energies where only rather few partial
waves contribute. This means that the unique constraint from cross sections is on the
asymptotic behavior of the potentials which however can have different
number of bound states and still precisely identical phase shifts.
This immediately implies that the scattering wave functions may differ at
short distances by a corresponding number of nodes. On the other hand, 
structure information is related to matrix elements of an operator between
initial and final states. As a consequence, structure information is
contained in the short-distance wave-function properties, and therefore
we could expect a substantial interaction dependence.

In this report we shall investigate these problems by use of
the simplest two-body process of inelastic $\alpha-\alpha$
scattering. This provides information about the continuum properties
of $^{8}$Be. We shall calculate cross sections defined by specified
initial and final state energies. We shall investigate dependence on
energies and on the chosen $\alpha-\alpha$ interaction.  We shall
extract as much structure information of $^{8}$Be as possible, and in
particular focus on ${\cal B}^{(E2)}$-values. However, these
quantities require initial and final states connected by the electric
quadrupole operator, and the appropriate continuum states are 
not a priory well defined.

Previous works provided cross sections from energies around the
positions of the $2^+$ and $4^+$ resonances in $^8$Be
\cite{lan86b,lan86c,kro87,dat05,hyl10,kir11,rog12}.  The theoretical
results \cite{lan86b,lan86c,kro87} are converging although important
discrepancies still remain. In these works the structure information is 
extracted as ${\cal B}^{(E2)}$-values by assuming a Breit-Wigner shape for
population of the decaying initial $2^+$ and $4^+$ resonances in
$^{8}$Be.  No continuum background contribution is considered.

The purpose of this work is to provide cross sections and ${\cal
  B}^{(E2)}$-values with the highest numerical accuracy, test the
dependence on how the inherent continuum state divergence is removed,
and investigate the effect of a different short-distance behavior of
the wave functions produced by the use of different potentials. More
precisely, we shall provide cross sections for energies covering all
positions of the resonances, $J^{\pi}=2^+,4^+,6^+,8^+$, and limited to
windows of final state energies around lower-lying resonances.  The
latter information is directly related to structure characteristics
and furthermore also the quantities measured in practice in
experiments. We shall compare to the scarce pieces of previous
theoretical and experimental results.

In Section II we describe the theoretical background needed for the
calculations.  The results are collected in Section III, which
contains three subsections describing, respectively, the $^8$Be
spectrum and its dependence on the two-body potentials, the electric
quadrupole cross sections for specific transitions, and the total
cross sections for different final energy windows. The extraction of the 
${\cal B}^{(E2)}$ transition strengths is described in Section IV. We finish 
in Section V with a short summary and the conclusions.

\section{Cross section expressions}

The inelastic two-body scattering process is described in detail in
Ref.~\cite{ald56} as a bremsstrahlung cross section (see also
Ref.~\cite{tan85}). In particular, the total cross section for $E\lambda$-radiation
is given by Eq.(9) of Ref.\cite{tan85}. Assuming a  two-body collision involving two 
identical particles with charge $Z$ and zero spin, the cross 
section becomes:
\begin{widetext}
\begin{equation}
\frac{d\sigma}{dE_\gamma} \bigg|_{\ell \rightarrow \ell^\prime}=
\frac{8\pi^2 Z^2e^2}{2^{2\lambda-2} k^2}
\frac{(\lambda+1) (2\lambda+1)}{\lambda[(2\lambda + 1)!!]^2}
(2\ell+1)^2 (2\ell^\prime+1)
\left(\frac{E_\gamma}{\hbar c} \right)^{2\lambda+1} 
\bigg| \langle \ell 0; \lambda 0| \ell^\prime 0\rangle
W(\lambda \ell^\prime \ell 0; \ell \ell^\prime)
\int_0^\infty u_\ell(E,r) r^{\lambda} u_{\ell^\prime}(E^\prime,r) dr
 \bigg|^2 \;,
\label{cross1} 
\end{equation}
\end{widetext}
where $E$ and $E^\prime$ are the initial and final energies in the
two-body center of mass frame, $E_\gamma=E-E^\prime$ is the energy of
the emitted photon, $\ell$ and $\ell^\prime$ are the relative angular
momenta between the two particles in the initial and final state, 
$k^2=2 \mu E/\hbar^2$ ($\mu$ is the reduced mass of the two-body
system), and $W$ represents a standard Racah coefficient.  

The radial two-body wave functions $u_\ell$ and $u_{\ell^\prime}$ are
solutions of the radial two-body Schr\"{o}dinger equation with
potential $V(r)$, and they obey the large-distance boundary condition
\begin{equation}
u_\ell(E,r) \stackrel{r \rightarrow \infty}{\longrightarrow} 
 \sqrt{\frac{2 \mu}{\pi \hbar^2 k}}
  \left[ \cos\delta_\ell F_\ell(kr) + \sin\delta_\ell G_\ell(kr)\right],
\label{asymp}
\end{equation}
where $F_\ell$ and $G_\ell$ are the regular and irregular Coulomb
functions, $\delta_\ell$ is the nuclear phase shift, and the normalization
constant is determined by the orthogonality condition:
\begin{equation}
\int_0^\infty u_\ell(E,r) u_\ell(E^\prime,r) dr=\delta(E-E^\prime).
\label{ener}
\end{equation}

The total bremsstrahlung cross section is finally obtained after integration
over the energy of the emitted photon:
\begin{equation}
 \sigma(E)=\int \left. 
 \frac{d\sigma}{dE_\gamma}\right|_{\ell \rightarrow \ell^\prime} 
 \hspace*{-5mm}(E) \;dE_\gamma  \;,
\label{intcs}
\end{equation}
where $E'=E-E_{\gamma}$.  In addition to the integration, a summation
over all angular momenta, $\ell$ and $\ell^{\prime}$, as well as all
multipolarities, $\lambda$, has to be included in Eq.(\ref{intcs}).
We shall avoid cluttering the notation by adding more indices. In any
case we also want to keep track of the individual contributions.

The measurement is simply to control relative energy of two colliding
particles, and identify the same outgoing particles and measure their
energies.  The computed cross sections should then be obtained in
close analogy to the experimental setup, where only a finite range of
final relative energies is measured. This means that the integral in
Eq.(\ref{intcs}) should be performed only over this precise energy
range. We shall often refer to this range as the final energy window.

Once a two-body potential has been chosen, the procedure is clear from
Eqs.(\ref{cross1}) and (\ref{intcs}), that is, find the wave functions
and perform the integrals.  However, some care must be exercised in
the calculation of the matrix elements because the continuum wave
functions, $u_{\ell}(r)$, cannot be normalized in coordinate space.
The bare matrix elements diverge until a regularization procedure is
employed. We shall follow the Zel'dovich prescription \cite{zel60},
which introduces the regularization factor $e^{-\eta^2 r^2}$ in the
radial integrand, such that the correct result is obtained in the
limit of zero value for the Zel'dovich parameter $\eta$.  Fortunately,
this removes the unwanted large-distance contributions and the
remaining physics results are uniquely defined, since they are stable
for sufficiently small values of $\eta$. Obviously, the smaller the
parameter $\eta$ the slower the fall off of the radial integrand, and
therefore the larger the upper limit required in the radial integral
in Eq.(\ref{cross1}).

In order to compute inelastic cross sections the only remaining decision 
is to choose precisely which observable should be computed, that is which 
energy interval has to be employed in the integration (\ref{intcs}). Except for practical 
experimental difficulties it is possible to choose any energy interval allowing
emission of one photon. The advantage is that no additional
information is required, as for example definition of a resonance or
knowledge of the decay mechanism through an intermediate structure. This information
is implicitly contained in the continuum wave functions. Note that when using
Eq.(\ref{asymp}) the energies $E$ and $E^\prime$ (and therefore
also $E_\gamma$) are treated as continuum variables (the continuum is not discretized). 
The integral (\ref{intcs}) can be easily performed within any arbitrary energy limits by 
choosing an also arbitrary small grid for the photon energy.

Direct comparison of precisely the same observables is then possible
and desirable.  However, an important question is whether the
structure information can be disentangled from the measured cross
sections.  This is a standard procedure for bound states, and the
obvious continuation is to apply the procedure to resonances. One
option is to focus on final energy windows around peaks in the cross
sections corresponding to (otherwise known) resonances.
Interpretation in terms of multipole transitions may then be possible
but may be also ambiguous if the results depend too much on the chosen 
energy windows.  It is crucial to know if the structure information can 
in principle be extracted.  In fact, most theoretical results are
obtained without even realizing this problem because the continuum
already is discretized and resonances therefore model dependently
defined.

Finally, it is also important to know the dependence of these observables on
the two-body potential. After all, different potentials providing
identical phase shifts could produce wave functions with a different
short-distance behavior, which therefore could also produce different
radial integrals in Eq.(\ref{cross1}) and different cross sections.

In the following section all these issues are investigated for the
$\alpha-\alpha$ scattering process.

\section{Computed cross sections}

The $\alpha-\alpha$ reaction is specified by initial and final state
energies, as well as by the energy window for measured final energies.
The angular momentum is at best only indirectly controlled in the
experiment.  However, knowledge of the two-body resonance properties
allow substantiated expectation of dominating individual angular
momenta at certain energy ranges. In particular, the cross section
is expected to have peaks for incident energies in the vicinity of the
$^{8}$Be resonances.  For the two-body $\alpha$-structure of $^{8}$Be
only even relative angular momenta and positive parity are allowed.
This means that the lowest multipolarity of an electromagnetic
transition must have $E2$ character. Therefore, the cross section is
given by Eq.(\ref{cross1}) for $Z=2$, $\lambda=2$, and where only even
values of $\ell$ and $\ell^\prime$ are possible. Transitions with
higher multipolarity, where $\lambda =4$ is the next allowed, are clearly
smaller by several orders of magnitude \cite{ald56,bla79}.

\subsection{Two-body potentials and $^8$Be spectrum}

\begin{table*}
\begin{center}
\caption{Resonance energies, $E_r$, and widths, $\Gamma$, in MeV  of the five lowest computed 
resonances in $^8$Be with angular momentum and parity $J^\pi$. Rows two and three give,
when available, the corresponding experimental values taken from Ref.\cite{til04}. 
The following four rows are the computed values with the Buck and Ali-Bodmer-d potentials
\cite{buc77,ali66}.  }
\label{tab1}
\begin{ruledtabular}
\begin{tabular}{|c|ccccc|}
  $J^\pi$  &  $0^+$  &   $2^+$  &    $4^+$  &    $6^+$  &    $8^+$  \\
\hline
$E_r$ (Exp.)        &         0.0918          & $2.94\pm0.01$   &   $11.35\pm0.15$      &  ---   &  ---      \\
$\Gamma$  (Exp.)  & $(5.57\pm0.25)\times 10^{-6}$ & $1.51\pm0.02$   &     $\sim 3.5  $      &  ---   &  ---      \\
\hline
 $E_r$ (Buck)       &          0.091           &      2.88       &      11.78            &  33.55  &  51.56   \\
 $\Gamma$ (Buck)  &   $3.6\times10^{-5}$      &      1.24       &       3.57            &  37.38  &  92.38   \\
\hline
 $E_r$  (Ali-Bodmer-d)   &          0.092           &      2.90       &      11.70            &  34.38  &  53.65   \\
$\Gamma$ (Ali-Bodmer-d)&   $3.1\times10^{-6}$      &      1.27       &       3.07            &  37.19  &  93.74   \\
\end{tabular}
\end{ruledtabular}
\end{center}
\end{table*}

Two different potentials will be used for the $\alpha-\alpha$
interaction: The Buck potential \cite{buc77} and the {\it d} version
of the Ali-Bodmer potential \cite{ali66}. Both of them are simple
gaussian potentials that fit equally well the low-energy ($E<15$ MeV)
$\alpha-\alpha$ phase shifts, not only for $\ell=0$, $\ell=2$, and
$\ell=4$ \cite{lan86b}, but also for $\ell=6$ and $\ell=8$. One of the
differences between both potentials is that while the Buck potential
is partial-wave independent, the Ali-Bodmer potential has been
adjusted separately for each individual partial wave.  The second and
most important difference is in the treatment of the Pauli
principle. The Buck potential contains two $s$-wave and one $p$-wave
Pauli-forbidden bound states \cite{sai69}, which in the case of the
Ali-Bodmer potential are automatically excluded by use of a repulsive
potential, also with gaussian shape. A third alternative is to use the
phase-equivalent version of the Buck potential, which is constructed
in such a way that the Pauli-forbidden states are removed from the
spectrum but keeping the phase shifts exactly the same.  Then also the two-body
resonances remain at exactly the same positions as for the original
potential \cite{gar99}.

In table~\ref{tab1} we give the energies ($E_r$) and widths ($\Gamma$)
of the lowest five resonances found in $^8$Be. They have been obtained
by use of the complex scaling method, which permits to extract
resonances as poles of the $S$-matrix and with the complex-scaled wave
functions behaving asymptotically as bound states \cite{ho83}. The
values in the second and third rows correspond to the experimentally
known resonances \cite{til04}.  As seen in the table, the two
potentials used in this work give rise to very similar spectra, not
only for the energies themselves, but also for the corresponding
widths. The computed widths for the experimentally unknown $6^+$ and
$8^+$ resonances are rather big, which means that even if they appear
as poles of the $S$-matrix they are pretty much diluted in the
two-body continuum.  They can hardly be characterized as resonances
related to observables. 

\subsection{Electric quadrupole cross sections}

\begin{figure}
\epsfig{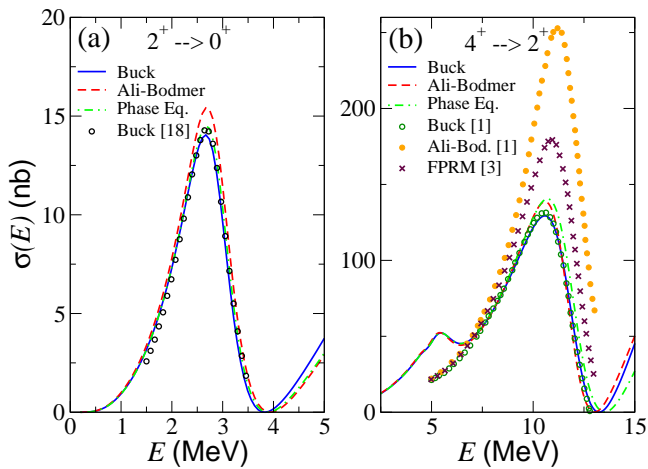}
\caption{(Color online) Integrated cross section (Eq.(\ref{intcs})) 
for the $2^+ \rightarrow 0^+$ (a) and  $4^+ \rightarrow 2^+$ (b) transitions 
as function of initial energy.  Results using the Buck potential (solid curves), 
the Ali-Bodmer-d potential (dashed curves), and the phase equivalent version of 
the Buck potential (dot-dashed curves) are shown. In (a), the open circles
are the result given in \cite{lan86}. In (b), the open and closed circles are the
results given in \cite{lan86b} obtained with the Buck and Ali-Bodmer potentials, 
respectively, and the crosses are the cross section calculated in \cite{kro87} using 
the Buck potential and the Finite Pauli Repulsion Model.}
\label{fig1}
\end{figure}

Let us start by investigating the dependence of the cross section on
the $\alpha-\alpha$ potential. As already mentioned, for
$\alpha-\alpha$ collisions the $E2$ contribution dominates. In
Fig.\ref{fig1} we show the computed electric quadrupole cross sections
for the $2^+ \rightarrow 0^+$ (left part) and $4^+ \rightarrow 2^+$
(right part) transitions. They have been obtained according to
Eqs.(\ref{cross1}) and (\ref{intcs}). For the $2^+ \rightarrow 0^+$
case, due to the extremely small width of the $0^+$ resonance (just a
few ev, see table~\ref{tab1}), the final energy window for the
integration in Eq.(\ref{intcs}) has been taken very small, of only 0.5
keV, which is far smaller than the best experimental resolution of
about $10$~keV.  For the $4^+ \rightarrow 2^+$ reaction we have chosen
the same final energy interval as in \cite{lan86b,lan86c,kro87},
namely, $2 \mbox{ MeV} < E^\prime < 4 \mbox{ MeV}$, which corresponds
roughly to the $2^+$ resonance energy $\pm 1$ MeV.

In the figure the solid, dashed, and dot-dashed curves correspond to
the cross sections obtained with the Buck potential \cite{buc77}, the
Ali-Bodmer-d potential \cite{ali66}, and the phase equivalent Buck
potential \cite{gar99}, respectively. As we can see, for both
transitions the computed cross sections are quite similar to each
other for the different potentials.  This similarity is also found in
spite of the fact that the nodes in the scattering wave functions are
different although the phase shifts are the same.  The surprise is
that the transition probability is determined by a matrix element
between these wave functions.  It could then easily have depended on
the nodes, but the contribution is apparently determined by the
identical structure at large distances.

Improvement in conceptual or numerical accuracy is
exhibited by comparison to deviating previous calculations. In particular, 
in Fig.~\ref{fig1}b the closed circles are the results reported in \cite{lan86b},
where the Ali-Bodmer potential was used, and the crosses correspond to
the cross section given in \cite{kro87}, where the Buck potential and
the Finite Pauli Repulsion Model was used. However, the calculations in \cite{lan86} (open circles
in Fig.\ref{fig1}a) and \cite{lan86b} (open circles in Fig.\ref{fig1}b)
corresponding to the $2^+ \rightarrow 0^+$ and $4^+ \rightarrow 2^+$ transitions,
respectively, and performed both with the Buck potential, agree very nicely
with our calculations with the same potential.

The cross sections in Fig.\ref{fig1} have been obtained with a Zel'dovich parameter
of $\eta=0.02$ fm$^{-1}$. For this value the computed cross sections are already stable
when going to the limit $\eta \rightarrow 0$. In principle, smaller values could have been used, but
the smaller $\eta$ the larger the distance up to which the integral in Eq.(\ref{cross1})
has to be performed. The increase of the integration distance makes the numerical calculation
of the integral more and more complicated due to the strongly  oscillating behavior of the
integrand. In any case, $\eta$ values up to twenty times smaller than the one used in 
Fig.\ref{fig1} produce cross sections that are indistinguishable from the ones shown in the
figure.

In all the cases the general behavior of the cross sections in Fig.\ref{fig1}  is qualitatively 
very similar.
The energy dependence exhibits a pronounced peak at about the $2^+$ resonance energy in
Fig.\ref{fig1}a, and about the $4^+$ resonance energy in Fig.\ref{fig1}b (see table~\ref{tab1}).
After decrease through a minimum the cross section increases at higher energies due to the
advantage of an increasing photon energy, see Eq.(\ref{cross1}), and a
fair match between some of the continuum states.  

In Fig.\ref{fig1}b a remarkable feature in the cross section appears at small energies.
The decrease of the integrated cross section with decreasing energy continues 
all the way to the threshold. However, the decrease is not smooth, and 
a small bump is observed. It is important to note that the bump appears precisely in 
the region where the initial energy $E$ approaches the upper limit of the energy window 
of final energies $E'$ ($\sim$4.0 MeV). This means the bump appears in the region where
the photon energy ($E_\gamma=E-E'$) takes small values. For the $2^+ \rightarrow 0^+$ reaction
(Fig.\ref{fig1}a) this bump is actually not seen due to the very small energy and width of the $0^+$
resonance and the very small size of the corresponding energy window.

\begin{figure}
\epsfig{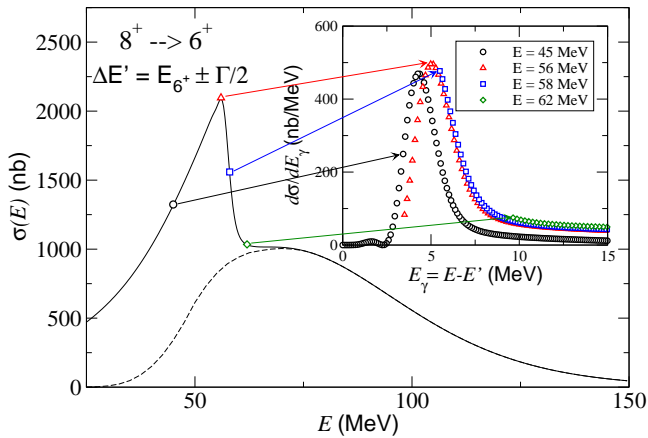}
\caption{(Color online) Outer part: Integrated cross section 
for the $8^+ \rightarrow 6^+$ transition in $^8$Be with a final energy
window of $E_{6^+}\pm \Gamma/2$. The solid and dashed curves are the calculation
including and excluding the soft-photon contributions, respectively.
Inner part: Differential cross section (integrand of Eq.(\ref{intcs})) as
a function of the photon energy $E_\gamma$ for the initial energy values
$E=45$ MeV (circles), $E=56$ MeV (triangles), $E=58$ MeV (squares),
and $E=62$ MeV (diamonds).  }
\label{fig2}
\end{figure}

To exhibit the origin of this bump, we shall consider the theoretical
transition of $8^+ \rightarrow 6^+$. According to table~\ref{tab1},
at least theoretically, the $8^+$ and $6^+$ resonances have a fairly
large width, with a substantial overlap between both resonances. In
the outer part of Fig.\ref{fig2} the solid line shows the $8^+
\rightarrow 6^+$ cross section computed following Eqs.(\ref{cross1})
and (\ref{intcs}) with the Buck potential. Only final energies within
the window $E_{6^+}\pm \Gamma/2$ have been considered ($E_{6^+}$ and
$\Gamma$ are the energy and width of the $6^+$ resonance).  This means
that the upper limit of the window is about 53 MeV. As we can see in
the figure, for energies below this value the cross section shows a
huge bump that does not match at all with the smooth behavior of the
cross section obtained for higher energies. This huge bump has the
same origin as the small bump observed in Fig.\ref{fig1}b.

To understand the origin of this bump it is very clarifying to look into the differential cross sections
for some specific values of the initial energy $E$ (integrand of Eq.(\ref{intcs}) for given values of
$E$). This is shown in the inner part of Fig.\ref{fig2}
as a function of the photon energy for initial energy values $E=45$ MeV (circles), $E=56$ MeV (triangles), 
$E=58$ MeV (squares), and 
$E=62$ MeV (diamonds).  Integration of the differential cross section as indicated in (\ref{intcs}) gives 
rise to the total cross section indicated in the outer part of the figure by the same symbols.

As we can see, for $E=62$ MeV, where the total cross section is still out of the bump, the differential
cross section shown in the inset is rather flat (diamonds), without any particular structure. 
The minimum value of $E_\gamma$ is of about 9 MeV, which is the difference between $E$ (62 MeV) and the upper
part of the energy window for $E^\prime$ (53 MeV).
However, for 
$E=58$ MeV a sharp increase appears in the differential cross section (squares). This increase becomes
a well defined and almost complete peak for $E=56$ MeV (triangles), which produces the maximum
of the bump in the cross section. For $E=45$ MeV (circles) the peak in the differential cross section
is now complete, but it is a bit smaller than for $E=56$ MeV, which gives rise to an also smaller
total cross section.

\begin{figure}
\epsfig{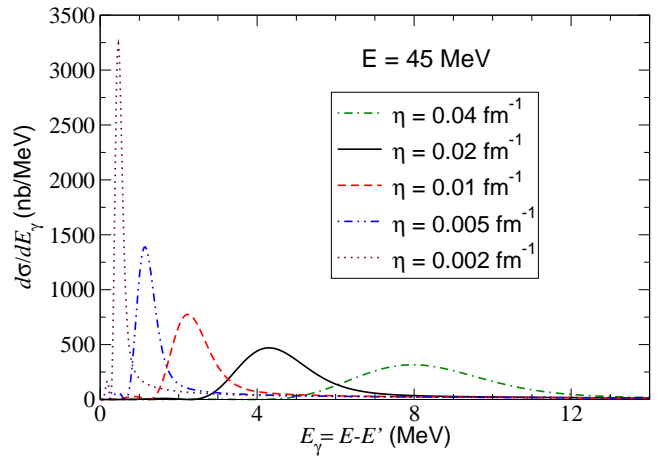}
\caption{(Color online) Differential cross section
(integrand of Eq.(\ref{intcs})) as a function of the photon energy $E_\gamma$
for the $8^+ \rightarrow 6^+$ transition in $^8$Be. The initial energy is $E=45$ MeV and a final 
energy window of $E_{6^+}\pm \Gamma/2$ has been used. The different curves correspond to different
values of the Zel'dovich parameter $\eta$ used to regularize the continuum wave functions.}
\label{fig3}
\end{figure}

The nature of this peak in the differential cross section is shown in
Fig.\ref{fig3}. In fact, although the total cross section does not
depend on the Zel'dovich parameter (provided that the value is small
enough), the differential cross section does show a dependence on the
Zel'dovich parameter. In Fig.\ref{fig3} we show the same differential
cross section as in the inset of Fig.\ref{fig2} for an initial energy
$E=45$ MeV and for different values of $\eta$. In all the cases the
integrated cross section is about the same, but as $\eta$ decreases
the peak of the differential cross section moves towards zero and
becomes sharper.  In fact, in the limit of very small $\eta$ values
the peak would be extremely sharp and located basically at
$E_\gamma=0$.

This behavior of the differential cross section is actually showing the known $1/E_\gamma$ dependence 
of the bremsstrahlung cross section at small photon energies \cite{gre01}. This is the so called
infrared catastrophe. This divergence
is behind the appearance of the bumps in the cross sections under discussion.
However, as explained in \cite{gre01}, this divergence is not physical. A transition with $E_\gamma=0$
is nothing but an elastic process. A relativistic treatment of the elastic reaction up to the
same order will produce a similar $1/E_\gamma$ divergence in the cross section but with opposite sign
that precisely cancels the one obtained in the calculation of the bremsstrahlung cross section.

A simple way of eliminating the unphysical contribution of the soft
photons is to exclude from the integral in Eq.(\ref{intcs}) the sharp
peak in the differential cross section shown in Fig.\ref{fig3}.  When
this is done, the resulting cross section for the $8^+ \rightarrow
6^+$ transition is shown by the dashed line in the outer part of
Fig.\ref{fig2}.  As we can see, exclusion of the soft photon
contribution eliminates the bump in the cross section.

\begin{figure}
\epsfig{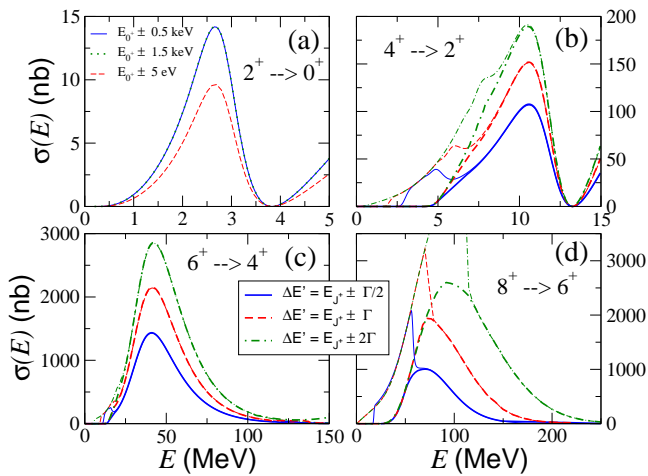}
\caption{(Color online) Integrated cross sections (Eq.(\ref{intcs})) as a function of the initial 
energy, $E$, for the $2^+ \rightarrow 0^+$, $4^+ \rightarrow 2^+$, $6^+ \rightarrow 4^+$, and
$8^+ \rightarrow 6^+$ transitions (panels (a), (b), (c), and (d), respectively). In panel (a)
the solid, dotted, and dashed curves correspond to final energy windows $E_{0^+}\pm 0.5$ keV,
$E_{0^+}\pm 1.5$ keV, and $E_{0^+}\pm 5$ eV, 
respectively, where $E_{0^+}$ is the energy of the $0^+$ resonance in $^8$Be. In panels (b), (c), and (d)
the solid, dashed, and dot-dashed curves correspond to final energy windows
$E_{J^\pi} \pm \Gamma/2$, $E_{J^\pi} \pm \Gamma$, and $E_{J^\pi} \pm 2\Gamma$, respectively,
where $E_{J^\pi}$ and $\Gamma$ are the
energy and width of the resonance in the final state with spin and parity $J^\pi$. The thick and
thin curves are the calculations with and without exclusion of the soft-photon contributions, respectively.}
\label{fig4}
\end{figure}

Finally, in Fig.\ref{fig4} we show how the cross sections for the
different transitions depend on the energy window chosen for the final
state. Due to the close similarity observed in Fig.\ref{fig1} for the
different potentials it suffices to show results for only one of
them, e.g. the Buck potential.

For the $2^+ \rightarrow 0^+$ transition, due to the very small width of
the $0^+$ resonance, an energy window of 0.5 keV around the resonance
energy already maximizes the cross section. In fact, when a window
three times wider ($E_{0^+}\pm 1.5$ keV) is used, both cross sections
are completely indistinguishable (solid and dotted curves in
Fig.\ref{fig4}a).  In order to observe some variation an extremely
narrow window should be considered.  As an example, the dashed curve
in the figure shows the cross section obtained with an energy window
of only $E_{0^+}\pm 5$ eV. In this case the maximum of the cross
section decreases from about 14 nb to a bit less than 10 nb.

In Figs.~\ref{fig4}b, \ref{fig4}c, and \ref{fig4}d the energy windows
$E_{J^\pi} \pm \Gamma/2$ (thick-solid curves), $E_{J^\pi} \pm \Gamma$
(thick-dashed curves), and $E_{J^\pi} \pm 2\Gamma$ (thick-dot-dashed
curves) have been used, where $E_{J^\pi}$ and $\Gamma$ are the energy
and width, respectively, of the resonance in the final state with spin
and parity $J^\pi$. The corresponding thin curves are the cross
sections obtained when the soft-photon contribution is included. This
results in the bump observed at small energies. As explained, the size
of the bump is directly related to the overlap between the final
energy window and the initial energy $E$. Even for the case of the
largest energy window, the effect of the soft photons is limited to
$E$-values below $\sim$6 MeV for the $4^+ \rightarrow 2^+$ transition and below
$\sim$18 MeV for the $6^+ \rightarrow 4^+$ transition. Only for the $8^+
\rightarrow 6^+$ transition, due to the large widths of the resonances
involved, this effect is clearly visible.

We notice in Figs. \ref{fig4}b, \ref{fig4}c, and \ref{fig4}d that the
increase of the size of the final energy window produces a significant
increase of the cross sections. This may be attributed to the
relatively broad final state resonances.  The variation can reach up
to a factor of 2 when changing from the smallest to the largest
window.  In comparison to measured values, this window dependence is
therefore very important.  It is tempting to discuss this window
dependence as arising by a division into resonance-to-resonance decay
and non-resonant continuum background contributions to the decay. Then,
increase of the window width to be larger than $\Gamma$ (full width at
half maximum) should entirely add only non-resonant contributions.
Such a distinction can at this level never be sharp and well defined,
although suggestive and probably even useful.  

We emphasize that the
window dependence so far has been only shown for initial and final state
energies around resonance positions and for the corresponding set of
angular momenta. In any case, these contributions are expected to be
the dominating ones, as shown in the next subsection.

\subsection{Total cross sections}

The total cross section corresponding to the $E2$-transition for a given final energy
window should contain not only the contributions shown in Fig.\ref{fig4}, which correspond to 
some specific transitions with given initial and final angular momenta, but they should also contain
the contributions from all the other possible $E2$ transitions to that precise final energy window.
In fact, the observables are first of all the cross sections restricted to be functions of the chosen 
initial and final state energies, and the contributions from different types of transitions can not be 
directly distinguished.
In particular, keeping aside the $8^+$ states, the $2^+ \rightarrow 0^+$, $0^+ \rightarrow 2^+$, 
$2^+ \rightarrow 2^+$, $4^+ \rightarrow 2^+$, $2^+ \rightarrow 4^+$, $4^+ \rightarrow 4^+$, 
$4^+ \rightarrow 6^+$, $6^+ \rightarrow 4^+$, and $6^+ \rightarrow 6^+$ transitions could contribute
to the total $E2$ cross section for a given final energy window.

To get a feeling on how these contributions can modify the cross sections shown in Fig.\ref{fig4},
let us focus first on the transitions between states with equal angular momentum and with final energy
windows around the resonance corresponding to that angular momentum. In other words, let us see how
the $2^+ \rightarrow 2^+$ transition contributes to the cross section in Fig.\ref{fig4}b, where the 
final energy window is chosen around the $2^+$ resonance energy, and the same for the contribution
of the $4^+ \rightarrow 4^+$ and $6^+ \rightarrow 6^+$ transitions to the cross sections
in Figs.\ref{fig4}c and \ref{fig4}d, respectively.

\begin{figure}
\epsfig{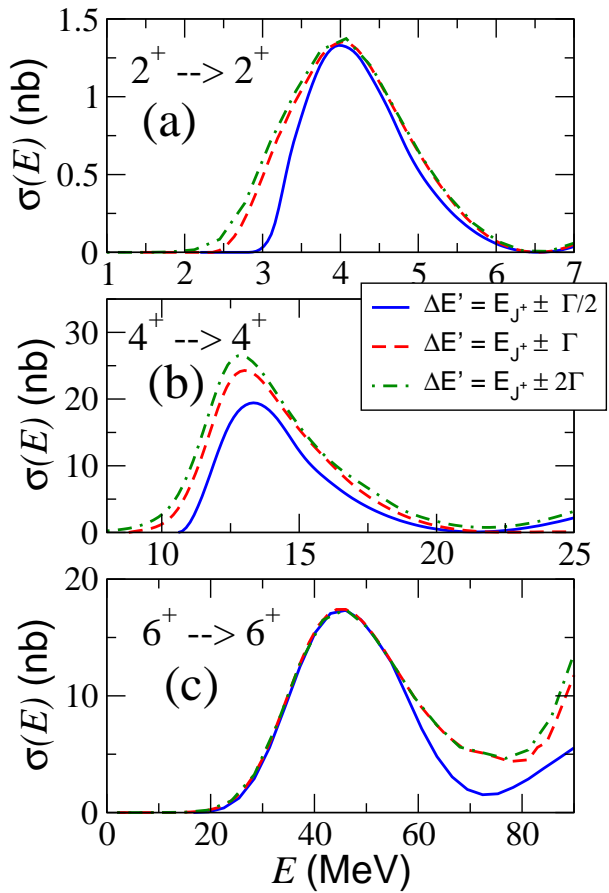}
\caption{(Color online) The same as in Fig.\ref{fig4} for the transitions 
$2^+ \rightarrow 2^+$, $4^+ \rightarrow 4^+$, and $6^+ \rightarrow 6^+$ 
(panels (a), (b), and (c), respectively). In all the calculations the 
soft-photon part has been excluded.}
\label{fig5}
\end{figure}

As in Fig.\ref{fig4}, the contribution from the $2^+ \rightarrow 2^+$, $4^+ \rightarrow 4^+$, and 
$6^+ \rightarrow 6^+$ transitions are expected to show a peak at an energy around one of the resonances
in the initial state, which in this case obviously coincides with the resonance in the final state. 
These transitions can then be understood as intrastate transitions for initial energies within the energy window around the resonances. It is then clear that in this case the removal of the soft-photon contribution 
becomes crucial. The corresponding cross sections for the $2^+ \rightarrow 2^+$, $4^+ \rightarrow 4^+$, 
and $6^+ \rightarrow 6^+$ transitions are shown in Fig.\ref{fig5}. The meaning of the different curves and 
the sizes of the final energy windows are the same as in Fig.\ref{fig4} for each of the final $J^+$ states.

The curves in Fig.\ref{fig5} have been computed for each of the
transitions by removing the contribution from the photon energies
corresponding to the unphysical peak in the differential cross section
analogous to the one shown in Fig.\ref{fig3}.  However, in this case,
the bulk of the cross section is affected by the soft-photon
contribution. Therefore, the computed cross section is very sensitive
to the cutoff imposed to the photon energy. In fact, in general, the
position of the peak to be removed in the differential cross section
depends on the initial energy, and the photon-energy cutoff should
then depend on this energy.  The curves in Fig.\ref{fig5} are then
estimates rather than accurate calculations. In any case, when
comparing to the curves in Fig.\ref{fig4} with the same $J^+$ final
state we observe that the contributions plotted in Fig.\ref{fig5} are
rather small, and they could produce an increase in the vicinity of
the maximum of the cross sections shown in Fig.\ref{fig4} of about 2
or 3\%.

Another interesting contribution to be analyzed in more detail is the one coming from transitions 
where the initial angular momentum is lower than the final one. They are the transitions $0^+ \rightarrow 2^+$,
$2^+ \rightarrow 4^+$, and $4^+ \rightarrow 6^+$. They are characterized by the fact that the 
resonance associated to the initial angular momentum is lower than the one associated to the
final angular momentum. When the final energy is limited to values within a window around 
the resonance, all the initial states will then be far from their
corresponding resonance, and they would be essentially pure continuum states.
As a consequence, the peak analogous to the one observed in Figs.\ref{fig4}b, \ref{fig4}c, and \ref{fig4}d,
located at an energy close to the one of the initial resonance, would not be there, and the 
contribution to the cross section is therefore expected to be not very relevant, and at most
enhance the tail of the cross section.

\begin{figure}
\epsfig{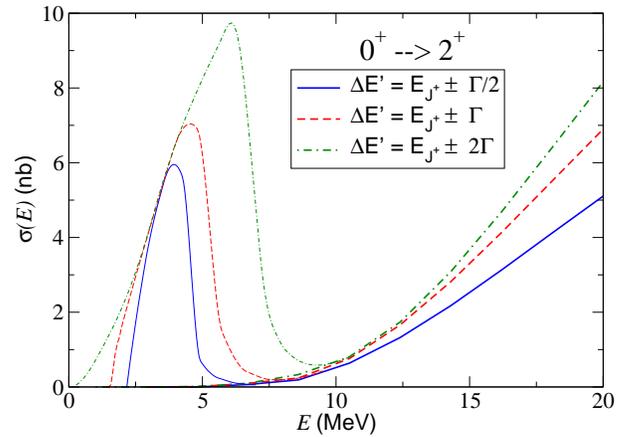}
\caption{(Color online) The same as in Fig.\ref{fig4} for the 
$0^+ \rightarrow 2^+$ transition. }
\label{fig6}
\end{figure}

As an illustration we show in Fig.\ref{fig6} the cross sections
corresponding to the $0^+ \rightarrow 2^+$ reaction for the three
usual final energy windows. The thick curves correspond to
calculations where the soft-photon contribution has been removed. The
computed cross sections are small in the energy region corresponding
to the peak in Fig.\ref{fig4}b, and their main contribution appears at
higher energies.  As in Fig.\ref{fig2}, the soft photons produce a
sizable bump (thin curves) when the initial energies are within the
energy window of the final state.  Obviously the wider the final
energy window the wider the energy range affected by the low-energy
photons, and therefore the wider the bump. How much of the bump
disappears after removal of the soft photons depends strongly on the
cutoff imposed to the photon energy. A cutoff estimated from the width
of the peak in the differential cross section (Fig.\ref{fig3}) makes
most of the bump disappear. In any case, even if the cross sections
shown in Fig.\ref{fig6} are a bit shaky in the energy region
corresponding to the bump, its weight is rather modest compared to the
cross sections shown in Fig.\ref{fig4}, at least for energies around
the maximum, and it amounts to no more than a few percent of the total.

\begin{figure}
\epsfig{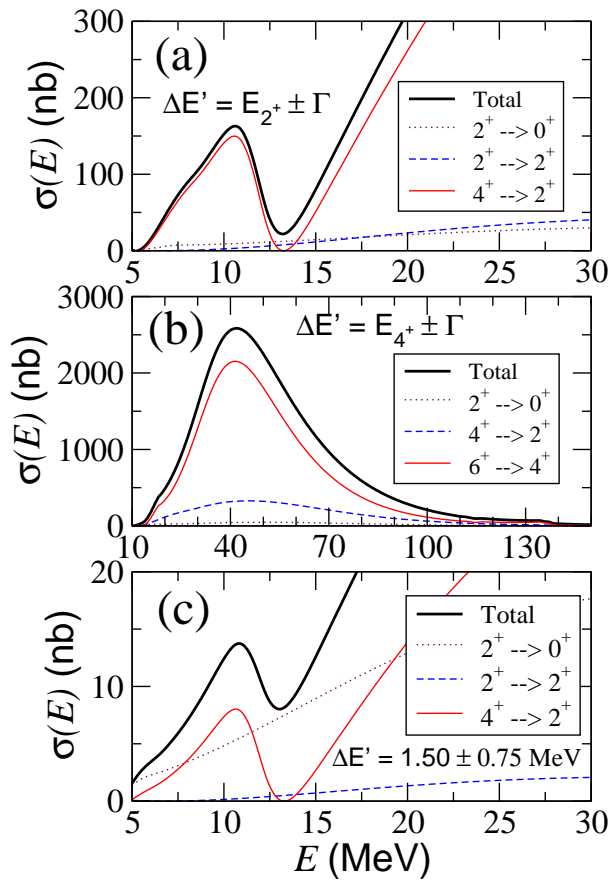}
\caption{(Color online) Total $E2$ cross sections for final energy 
windows (a) $E_{2^+} \pm \Gamma$, (b) $E_{4^+} \pm \Gamma$ , and 
(c) $1.75 \pm 0.75$ MeV. The total cross sections are given by the thick solid
curves. Although all the possible transitions are included in the
total, only the three of them giving the largest contributions are shown on
each of the three panels.}
\label{fig7}
\end{figure}

Finally, in Fig.\ref{fig7} we show the total cross section for three
different final energy windows. Together with the total, 
the three most relevant contributions among all the possible contributing
transitions are also shown.  In the upper and middle panels 
(Figs.\ref{fig7}a and \ref{fig7}b) the windows have been chosen around 
the $2^+$ and the $4^+$ resonance energies ($\pm \Gamma$), respectively.
In the lower part, Fig.\ref{fig7}c,
the final energy window has been taken in such a way it does not overlap
with any of the known resonances in $^8$Be. In particular, we have chosen
the energy window $1.5\pm0.75$ MeV.

Fig.\ref{fig7}a shows the expected peak from the leading
resonance-to-resonance transition.  The corresponding contribution was
shown by the thick-dashed curve in Fig.\ref{fig4}b, and it is now
given by the thin-solid curve.  This $4^+ \rightarrow 2^+$
contribution dominates both at the resonance and beyond. The increase
with energy is dramatic from zero just above the $4^+$ resonance
energy to values much larger than at the resonance.  This increase of
the non-resonant contribution is much less than corresponding to the
fifth power of the photon energy, which therefore reveals that the
matrix elements are decreasing.

Only a few of the other possible non-resonant transitions produce a
visible contribution to the cross section.  These are the ones shown
in the figure by the thin-dotted, thin-dashed, and thin-dot-dashed
curves, which correspond, respectively, to the $2^+ \rightarrow 0^+$,
$2^+ \rightarrow 2^+$, and $0^+ \rightarrow 2^+$ transitions.  They
represent the smooth non-resonant background contributions which away
from resonances (both final and initial energies) could be dominating.
In any case, although visible, their weight is not very significant,
and the total cross section (thick solid curve) follows the trend
dictated by the resonance-to-resonance transition.  At the resonance
peak they increase the maximum of the cross section from about 150 nb
to about 165 nb.

The cross section shown in Fig.\ref{fig7}b shows a similar structure
to the one in the upper panel. Now the resonance-to-resonance
contribution correspond to the $6^+ \rightarrow 4^+$ (thick-dashed
curve in Fig.\ref{fig4}c), and it is now given by the thin-solid
curve. Among the remaining transitions only the $4^+ \rightarrow 2^+$
gives a sizable contribution (dashed curve), which raises the maximum
of the total cross section from about 2150 nb to about 2550 nb.

The last part of the figure, Fig.\ref{fig7}c, corresponds to a
final-energy window whose center does not match with any of the known
resonances in $^8$Be. In particular, we have chosen $E'$ within the
window $1.50\pm 0.75$ MeV, which at most overlaps with the low-energy
tail of the $2^+$ resonance.  However, the cross section still
maintains some similarity with that of the $4^+$ resonance structure,
and a bump similar to the one in Fig.\ref{fig7}a, originating from the
$4^+ \rightarrow 2^+$ transition, appears again, although at a clearly
smaller scale.  As the energy increases, this contribution increases
dramatically as in Fig.\ref{fig7}a, but now there is no resonance
connection neither in initial nor in final state.  The remaining
contributions in Fig.\ref{fig7}c produce a structureless background,
among which the $2^+ \rightarrow 0^+$ is particularly big.  It even
becomes dominating above $20$~MeV.  This smooth behavior is typical
for transitions where both initial and final states are without
relation to any of the resonances.

The remaining contributions in Fig.\ref{fig7}c produce a structureless
background, among which the $2^+ \rightarrow 0^+$ is particularly
big. This smooth behavior is typical for transitions where both
initial and final states are without relation to any of the
resonances.  In any case, the total cross section (thick-solid curve)
still shows the bump around 11 MeV produced by the $4^+$ resonance.

\section{${\cal B}^{(E2)}$ transition strengths}

The computed cross sections are directly observables which by definition
contain both dynamics and structure information.  It is then
desirable to extract the transition probabilities, only
related to the structure,  which in turn are related
to the transition strengths ${\cal B}^{(E\lambda)}$-values \cite{lan86,lan86b}. 
In this work two different methods will be used to extract such strengths.

The first one requires assumptions about population of the resonances in the 
reaction. Once the resonance is populated, the cross section $\sigma(E)$ for the decay 
should approach a Breit-Wigner shape around the maximum \cite{bohr2}, which should be found 
at an energy $E$ close to the resonance energy. This is the method used in 
\cite{lan86,lan86b,kro87}. Clearly this assumption is only correct in the very nearest neighborhood
of the peak.  This furthermore assumes that no significant additional
smooth background contribution is present below the resonance peak.

Therefore, in the vicinity of the resonance, and for the particular case in which two $\alpha$-particles  
populate a resonance with angular momentum $J$, the cross section can approximated by:
\begin{equation}
\sigma(E)=(2J+1)\frac{\pi}{\kappa^2}\frac{\Gamma(E)\Gamma_\gamma}{(E-E_J)^2+0.25 \Gamma(E)^2},
\label{eq5}
\end{equation}
where $\kappa^2=2 \mu_{\alpha\alpha}E/\hbar^2$, $E_J$ is the energy of the resonance, and
\begin{equation}
\Gamma(E)=\Gamma_J \exp(2b/\sqrt{E_J}-2b/\sqrt{E}) 
\label{eq6} 
\end{equation}
with $b=2\pi e^2\sqrt{2\mu_{\alpha\alpha}/\hbar^2}$, and $\Gamma_J$ the computed (or the experimental)
resonance width \cite{gar11}.

If the cross section is assumed to have the form (\ref{eq5}), the $\Gamma_\gamma$ value can then be
 fitted to the maximum of the computed cross section, and using that \cite{asj}
\begin{equation}
\Gamma_\gamma= \frac{8\pi(\lambda+1)}{\lambda[(2\lambda+1)!!]^2}
\left( \frac{E_\gamma}{\hbar c} \right)^{2\lambda+1} {\cal B}^{(E\lambda)}(J\rightarrow J^\prime),
\label{eq7}
\end{equation}
the value of ${\cal B}^{(E\lambda)}$ can be immediately extracted. 

The transition strength above is the one
corresponding to the photoemission process $J \rightarrow J^\prime$, where $J^\prime$ is the
angular momentum of the final state with lower energy. The ${\cal B}^{(E\lambda)}$ value for a given 
transition and the inverse one are related by the simple expression:
\begin{equation}
{\cal B}^{(E\lambda)} (J \rightarrow J^\prime)  =\frac{2J^\prime+1}{2J+1}
{\cal B}^{(E\lambda)} (J^\prime \rightarrow J).
\end{equation}

\begin{table}
\caption{Widths for $\gamma$-decay ($\Gamma_\gamma$ in eV) and ${\cal B}^{(E2)}$-values 
(in $e^2 \mbox{fm$^4$}$) for the transitions given in the first
column. The labels `wide', `medium', and `narrow' refer to the size of the three
windows used in Fig.\ref{fig4} for each reaction.  The transition strengths obtained from 
$\Gamma_\gamma$ and Eq.(\ref{eq7}) are denoted by ${\cal B}_\gamma^{(E2)}$.
The transition strengths obtained by integrating under the peaks in the cross sections 
in Fig.\ref{fig4} (Eq.(\ref{eq9})) are denoted by ${\cal B}_\sigma^{(E2)}$.
When available, results from other works are given in the rows $\Gamma_\gamma^{\mbox{\scriptsize other}}$
and  ${\cal B}_{\mbox{\scriptsize other}}^{(E2)}$ .}
\label{tab2}
\begin{center}
\begin{tabular}{|c|c|ccc|}
\hline
   &   &   Wide  &  Medium  &  Narrow \\ \hline
$2^+ \rightarrow 0^+$  &  $\Gamma_\gamma$   & $7.7 \times 10^{-3}$ & $7.7 \times 10^{-3}$  & $5.2 \times 10^{-3}$ \\
&  $\Gamma_\gamma^{\mbox{\scriptsize other}}$& $8.3 \times 10^{-3 (a)}$   & $8.3 \times 10^{-3 (a)}$    &   \\  
                       &  ${\cal B}_\gamma^{(E2)}$  &  79.1  &  79.1   &  53.4 \\
                       &  ${\cal B}_\sigma^{(E2)}$  &  48.4  &  48.4   &  32.9 \\
&  ${\cal B}_{\mbox{\scriptsize other}}^{(E2)}$   &  $71^{(a)}$, $14.8^{(b)}$  & $71^{(a)}$, $14.8^{(b)}$  &  \\  \hline
$4^+ \rightarrow 2^+$  &  $\Gamma_\gamma$        &  0.59  & 0.47    & 0.33  \\
&  $\Gamma_\gamma^{\mbox{\scriptsize other}}$&    &  0.45$^{(c)}$, 0.46$^{(c)}$   &   \\  
                       &  ${\cal B}_\gamma^{(E2)}$  &  27.7  & 22.1    & 15.5  \\
                       &  ${\cal B}_\sigma^{(E2)}$  &  21.6  & 17.2    & 12.1  \\
&  ${\cal B}_{\mbox{\scriptsize other}}^{(E2)}$&    &  $18^{(c)}$,18.2$^{(b)}$, $25\pm 8^{(d)}$    &   \\  \hline
$6^+ \rightarrow 4^+$  &  $\Gamma_\gamma$        &  247   & 187    & 123  \\
                       &  ${\cal B}_\gamma^{(E2)}$  &  13.4  & 10.1   & 6.7  \\
                       &  ${\cal B}_\sigma^{(E2)}$  &  9.6  &   6.9   & 4.5  \\ \hline
$8^+ \rightarrow 6^+$  &  $\Gamma_\gamma$           & 832   &  633   &  323 \\
                       &  ${\cal B}_\gamma^{(E2)}$  & 17.1  &  13.0  &  6.6  \\
                       &  ${\cal B}_\sigma^{(E2)}$  & 9.9 &  5.2   &  2.5  \\ \hline

\end{tabular}
\end{center}
(a) Ref.\cite{lan86}, (b) Ref.\cite{wir00}, (c) Ref.\cite{lan86c,kro87}, (d) Ref.\cite{dat05}
\end{table}

Following the procedure just described, and using the computed cross sections shown in Fig.\ref{fig4},
we obtain the $\Gamma_\gamma$-values given in Table~\ref{tab2} for the different reactions
under investigation.  In the table the labels
`wide', `medium', and `narrow' refer to the sizes of the three final energy windows used in 
Fig.\ref{fig4} for each reaction. As a general rule, of course, the smaller the size of the window
the lower the maximum in the cross section, and therefore the smaller the value of $\Gamma_\gamma$. 
The only exception is the $2^+\rightarrow 0^+$ reaction, where the `wide' and `medium' windows produce 
the same cross section and therefore also the same $\gamma$-width.

In \cite{lan86}, a value of $\Gamma_\gamma=8.3$ meV is given for the $2^+\rightarrow 0^+$ reaction.
This value is consistent with the 7.7 meV obtained in our calculation, as expected due to the good
agreement obtained between our cross section and the one given in \cite{lan86} (open circles and
solid curve in Fig.\ref{fig1}a). The same happens with the $4^+\rightarrow 2^+$.  The cross section
obtained in \cite{lan86b} agrees well with our calculation (open circles and solid curve in 
Fig.\ref{fig1}b), whose maximum is very similar to the one shown in Fig.\ref{fig4}b for the
`medium'-size energy window ($E_{2^+}\pm \Gamma$). For this reason the $\Gamma_\gamma=0.47$ eV obtained in
our calculation for this particular case agrees well with the 0.46 eV given in \cite{lan86c}.
In \cite{kro87} also a very similar value of 0.45 eV is given.

After having computed the $\Gamma_\gamma$ values, Eq.(\ref{eq7}) permits us to obtain the transition strength. 
However, due to the $2\lambda+1$ exponent, the value of ${\cal B}^{(E\lambda)}$ is very sensitive to the 
value of $E_\gamma$ used. Already for $\lambda=2$ a variation of about 4\% in $E_\gamma$  amounts to an about 
20\% variation in the extracted transition strength. In our calculations the photon energy
($E_\gamma=E-E^\prime$) has been 
taken with $E$ equal to the energy at which the cross sections in Fig.\ref{fig4} have a maximum
(2.7 MeV in (a), 10.6 MeV in (b), 41.0 MeV in (c), and 70 MeV in (d)), and $E^\prime$ is taken 
to be the resonance energy in the final state. When this is done we obtain the values denoted in
Table~\ref{tab2} as ${\cal B}_\gamma^{(E2)}$.

Again, for the two largest energy windows in the $2^+\rightarrow 0^+$ reaction, the good agreement
with the $\Gamma_\gamma$ value given in \cite{lan86} implies also a good agreement with the 
${\cal B}_\gamma^{(E2)}$ value given in the same reference (75 W.u. $\approx 71$ $e^2$fm$^4$).
Note that our computed value is larger than the one obtained in \cite{lan86} in spite of the fact that
our computed $\Gamma_\gamma$ is smaller. This seems to be inconsistent with Eq.(\ref{eq7}) which 
implies that a smaller $\Gamma_\gamma$ should produce also a smaller ${\cal B}_\gamma^{(E2)}$. The reason
is the large dependence on $E_\gamma$ mentioned above. The values of $\Gamma_\gamma$ and 
${\cal B}_\gamma^{(E2)}$ given in \cite{lan86} are consistent with $E_\gamma \approx 2.7$ MeV, while
we have used a value of 2.6 MeV. For $\Gamma_\gamma=7.7$ meV, use of $E_\gamma=2.7$ MeV or $E_\gamma=2.8$ MeV
would reduce the ${\cal B}_\gamma^{(E2)}$ value down to 65.5 $e^2\mbox{fm$^4$}$ or 
54.6 $e^2\mbox{fm$^4$}$, respectively. Due to the smaller relative value of $E_\gamma$
compared to the other reactions, the $2^+\rightarrow 0^+$ reaction is more sensitive
to small variations in $E_\gamma$ than the other cases.
Another remarkable fact is that
our result (and the one in \cite{lan86}) clearly disagrees with the 14.8 $e^2$fm$^4$
given in \cite{wir00} where a Quantum Monte Carlo calculation is performed.

For the $4^+\rightarrow 2^+$ reaction the ${\cal B}_\gamma^{(E2)}$ value of 19 W.u. ($\approx 18$ $e^2$fm$^4$)
given in \cite{lan86c} agrees well with our calculation (as expected again due the good
agreement in the $\Gamma_\gamma$ value). In \cite{kro87} the same value of 19 W.u. is given. 
In our calculation a value of $E_\gamma\approx 7.7$ MeV has been used. As discussed above, a small variation
of $E_\gamma$ is not as relevant as in the $2^+\rightarrow 0^+$ case. For example, in the
`medium' window case, $E_\gamma =7.8$ MeV or $E_\gamma =7.9$ MeV gives rise to 
${\cal B}_\gamma^{(E2)}=20.7$ $e^2\mbox{fm$^4$}$ or
19.4 $e^2\mbox{fm$^4$}$, still pretty similar to the value given in table~\ref{tab2}.
It is surprising that the Quantum Monte 
Carlo calculation shown in \cite{wir00}, contrary to what happened for the $2^+\rightarrow 0^+$ 
reaction, agrees now very well with our result, since in \cite{wir00} a value of 
18.2 $e^2$fm$^4$ is given. More recently, an experimental value of $25\pm 8$ $e^2$fm$^4$
has been given in \cite{dat05} for the ${\cal B}_\gamma^{(E2)}$ transition strength
for the $4^+\rightarrow 2^+$. This value agrees as well with our results.

The second method that we use to obtain the transition strengths follows from the well 
known relation between the cross section $\sigma(E)$ in Eq.(\ref{intcs}) and
$d{\cal B}^{(E\lambda)}/dE$. In particular, this relation reads \cite{for03}:
\begin{eqnarray}
\lefteqn{ \hspace*{-1.5cm}
\sigma^{(E\lambda)}(E)=\frac{(2 \pi)^3(\lambda+1)}{\lambda \left[ (2\lambda+1)!!\right]^2}
\frac{1}{k^2} \left( \frac{E_\gamma}{\hbar c} \right)^{2\lambda+1} 
} \nonumber \\ && \times
\frac{2(2J+1)}{(2J_a+1)(2J_b+1)} \frac{d{\cal B}^{(E\lambda)}}{dE}(J\rightarrow J^\prime),
\label{eq9}
\end{eqnarray}
where $J_a$ and $J_b$ are the angular momenta of the two colliding particles ($J_a=J_b=0$
in our case) and $k^2=2\mu_{ab}E/\hbar^2$ ($\mu_{ab}$ is the reduced mass).

We can then choose to define another suitable value of ${\cal
  B}_\gamma^{(E\lambda)}$, that is by integration of
$\sigma^{(E\lambda)}(E)$ over the energy $E$, followed by dividing the
result by the factors multiplying $d{\cal B}^{(E\lambda)}/dE$ in
Eq.(\ref{eq9}). In other words, ${\cal B}_\gamma^{(E\lambda)} \propto \int
\sigma^{(E\lambda)}(E) / \langle E_\gamma\rangle^{2\lambda+1}$.  Here the choice of
the average value of 
$E_\gamma$ is essential due to the $2\lambda+1$ exponent, as discussed
when extracting ${\cal B}_\gamma^{(E\lambda)}$ from Eq.(\ref{eq7}).
We proceed also in this case as done in Eq.(\ref{eq7}), i.e., we take
this average value to be
$E_\gamma=E-E^\prime$ where $E$ is the energy at which the cross
sections in Fig.\ref{fig4} have a maximum and $E^\prime$ is the
resonance energy in the final state.

The results obtained for the cases shown in Fig.\ref{fig4} are denoted in Table~\ref{tab2} 
by ${\cal B}_\sigma^{(E2)}$. The integration over $E$ of the cross sections in Fig.\ref{fig4} are made
from 0 to the energy of the first zero (3.8 MeV, 13.3 MeV, 125 MeV, and 250 MeV in (a), (b), (c), and
(d), respectively).

Having in mind the large dependence of the computed transition strengths on the photon energy and 
the two completely different approaches used, the agreement between the ${\cal B}_\gamma^{(E2)}$ and the
${\cal B}_\sigma^{(E2)}$ values given in table~\ref{tab2} are quite reasonable. In all the cases 
${\cal B}_\sigma^{(E2)}$ is smaller than ${\cal B}_\gamma^{(E2)}$, especially for the 
$8^+ \rightarrow 6^+$ reaction, where the difference is essentially of a factor of 2.
The best agreement between both calculations is found for the two cases where the resonances
in $^8$Be are well established, namely, the $2^+\rightarrow 0^+$ and the $4^+\rightarrow 2^+$
reactions (keep in mind that the $2^+\rightarrow 0^+$ is particularly sensitive to $E_\gamma$).

\section{Summary and Conclusions}

We investigate electromagnetic transitions between continuum states.
The corresponding bound state problem is very well established and the
extension to the continuum should be straightforward.  This seems not
to be controversial, since the resonances are both the most prominent
structures in the continuum and the natural extension of the series of
discrete bound states.  However, in between resonances are well
defined continuum states, and the resonances themselves are
distributed over energy intervals in accordance with their widths.
Thus, at least at first glance the extension to the continuum is not
well defined.

In spite of these theoretical reservations, an increasing amount of
experimental activities are documented by a number of recent
publications.  The demand for theoretical understanding and
interpretation of measured continuum properties are therefore
increasing. The simplest non-trivial problem containing information
about continuum structure is inelastic scattering of two particles.
Already three particles present additional difficulties. Since the
two-body problem is technically much simpler, and still exhibits the
generic characteristics of continuum properties, we choose to
illustrate the concepts by $\alpha-\alpha$ scattering using
well-tested realistic potentials.

We first present the expression for the inelastic scattering cross
section as function of initial and final state energies. This
implicitly defines the energy of a necessarily emitted photon, which
does not have to be detected if particles and energies of both
initial and final states are precisely known.  The bosonic nature of
the $\alpha$-article limits the relative angular momenta to be even and
the parities to be positive.  Then the lowest multipolarity of the
emitted photon is $2^+$, but $4^+$, $6^+$ are allowed as well.  The
cross section expression is deceivingly simple and, except for the
kinematic factors, similar to the corresponding bound state transition
probability.  However, one divergence must be removed before
meaningful results can be obtained, that is the unnormalizable
resonance wave functions must be regularized.

The numerical calculations are then performed in close agreement with
a possible experimental procedure, that is select an appropriate
interval of final state energy and compute the cross section as
function of initial energy.  We first compare results from different
phase-shift identical potentials and from the few existing previous
computations of specific transitions.  The results from different
potentials are virtually indistinguishable, while previous results
($4^+ \rightarrow 2^+$) deviate substantially for an initial energy
around that of the $4^+$ state.

The chosen energy interval is the window where a perceived experiment
would measure energies of the outgoing $\alpha$-particles.  The
dominating contributions come, as intuitively expected, from
transitions between resonance states. The cross section is now
computed as a function of initial energy for photon emissions into an
energy window of width comparable to the width of the final state
resonance.  When the initial energy is within the energy interval,
allowing zero photon energy, an unphysical bump appears.  This
infrared catastrophe is due to omission of the elastic scattering
channel.  We remove the bump by omitting a corresponding peak,
low-lying and well-defined, corresponding to inclusion of large
distances before regularizing in the calculation of matrix elements.

Increasing the size of the energy window obviously increases the cross
section.  If the transition would be entirely from resonance to
resonance, we should find convergence with energy window
size. However, this is only seen for $2^+ \rightarrow 0^+$, while the
other cases keep increasing with inclusion of states beyond the width
of the resonance.  The largest contributions are by far for initial
energy within a resonance width, but even then non-resonant
contributions must be responsible for the continued increase of the
cross section.  This is part of a smooth and significant continuous
background, and variation of window size allows observable distinction
between resonance and background contributions.

The general size of the cross section increases rather strongly with
initial energy, but the largest resonance to resonance contributions
at the same time correspond to larger photon energies. However, the
actual increase is rather an order of magnitude smaller than dictated
by the fifth order dependence of $E2$-transitions.

It is remarkable that an initial energy close to the established $2^+$
and $4^+$ resonance positions plus half their widths produce vanishing
cross sections, independent of window size, for the dominating
$E2$-transitions.  This reflects destructive interference between
background and resonance amplitudes Further increase of the energy
above the resonance peaks leads again to rising cross sections.  For
sufficiently separated resonance peaks such an increase could produce
dominating contribution between the resonances and a significant
contribution within the next resonance peak.  In both cases this
amounts to pieces of non-resonant transitions between continuum
states.

Several transitions are allowed by angular momentum rules without
corresponding to resonance-to-resonance transitions. This is already
suggested by the previously discussed increase above the resonances,
but a number of other transitions may also contribute.  These are
first of all $E2$-transitions between continuum states of the same
angular momenta, which have been called intra-band transitions although
they have nothing to do with bands.  Also ``reverse'' transition, $J
\rightarrow J+2$, contribute.  These smaller transitions cannot in
principle be experimentally singled out, but our estimates show they
have to be included on the $2-10$~\% level in theoretical comparison.
One way to emphasize these terms is by using the theoretical guidance
to focus on initial energies between resonances, where their
contributions are comparatively larger.

Traditionally, structure information has been deduced from scattering
and reaction experiments.  This is straightforward for bound-to-bound
state transitions where photon energy is well defined, and strength
functions are simply related to decay rates and to cross sections.
However, for continuum-to-continuum transitions the relation is more
complicated, even though the same ingredients enter as matrix elements
and photon energy.  First the problem must be defined as related to
resonance properties.  This immediately emphasizes the ambiguity since
resonances only can be seen in observable quantities as peaks in cross
sections.  A resonance state is not defined as a state with specific
properties.

We show that extraction of both electromagnetic decay rates and
transition probabilities is inherently ambiguous but possible to
define and subsequently estimate with some uncertainty related to the
chosen definition. We use two definitions, that is the first where a
Breit-Wigner cross section is fitted very close to the resonance peak
energy resulting in a normalization closely related to the dominating
decay rate. The transition probability is then deduced from the
bound-to-bound state expression with a photon energy defined as the
difference between resonance energies. The second method is first to
integrate the cross section, limited to the corresponding transition,
over the initial energy from zero and across the resonance until its
is zero.  Subsequently we do as for the first method, i.e.  define the
appropriate photon energy and use the same relations to transition
probability.

The two methods turn out to give comparable results for the
established $2^+$ and $4^+$ resonances. The results are consistent
with the only measured value for the $4^+ \rightarrow 2^+$ transition,
and in agreement with one of the two previous cluster model
calculations.  Bound state approximations of resonances already
discretized the continuum and the few existing results show agreement
for the $4^+ \rightarrow 2^+$ transition, but rather large discrepancy
for the $2^+ \rightarrow 0^+$ transition.

In summary, we have investigated the electromagnetic continuum
transitions by computing two-body inelastic scattering cross sections.
We discuss results as functions of energies and angular momenta of
initial and final states. This includes the dominating
resonance-to-resonance contributions as well as other combinations
contributing to non-resonance smooth background decays.  Definitions
of decay rates and transition probabilities are shown to be inherently
ambiguous. We present two simple and intuitively appealing
possibilities where structure information is extracted.  The
experimental techniques are rapidly refined and a variety of details
can be expected in the near future.  The perspective and the interest
is now to extend these investigations beyond the two-body level. This
means first of all formulation and calculation for three-body systems
in the continuum.

\acknowledgments This work was partly supported by funds provided by
DGI of MINECO (Spain) under contract No. FIS2011-23565.  We appreciate
valuable continuous discussions with Drs. H. Fynbo and K. Riisager.

\end{document}